\documentclass[a4paper,10pt]{article}
\usepackage[latin1]{inputenc}
\usepackage[italian, english]{babel}
\usepackage{graphicx}
\usepackage{latexsym}
\usepackage{amssymb}
\usepackage{amsmath}
\usepackage{amsfonts}
\usepackage[margin=2cm]{geometry}
\usepackage[usenames,dvipsnames]{color}
\usepackage{setspace}
\usepackage[strict]{changepage}
\usepackage{booktabs}
\usepackage{float}
\usepackage{epsfig} 
\usepackage{graphicx, rotating}
\usepackage{hyperref}
\usepackage{slashed}
\usepackage{ifpdf}
\usepackage{cite}
\usepackage{multicol}
\usepackage{color}

\hypersetup{colorlinks, citecolor=bluscuro2, linkcolor=black, urlcolor=bluscuro2}
\definecolor{rossos}{rgb}{0.8,0.2,0.3}
\definecolor{bluscuro1}{rgb}{0.2, 0.2, 0.7}
\definecolor{bluscuro2}{rgb}{0.15, 0.2, 0.9}
\definecolor{verdes}{rgb}{0.1, 0.5, 0.1}

\newcommand{\be}{\begin{eqnarray}}
\newcommand{\ee}{\end{eqnarray}}
\newcommand{\bi}{\begin{itemize}}
\newcommand{\ei}{\end{itemize}}

\newcommand{\bea}{\begin{eqnarray}}
\newcommand{\eea}{\end{eqnarray}}

\newcommand{\vx}{\vec{x}} 
\newcommand{\vk}{\vec{k}}

\newcommand{\vq}{\vec{q}}
\newcommand{\fnl}{f_{\rm NL}}

\newcommand{\tnl}{\tau_{\rm NL}}

\begin{document}

\color{black}

\vspace{5truecm}
\begin{center}
{\Large\bf\color{black} Testing the Running of non-Gaussianity \\ 
\vspace{0.3cm} through the CMB $\mu$-distortion and the Halo Bias}\\
\bigskip\color{black}\vspace{1cm}{
{\large Matteo Biagetti, Hideki Perrier, Antonio Riotto and Vincent Desjacques
\vspace{0.8cm}
} \\[7mm]
{\em  
Universit\'e de Gen\`eve, Department of Theoretical Physics and Center for Astroparticle Physics (CAP),\\ 24 quai E. Ansermet, CH-1211 Geneva 4, Switzerland}}\\
\end{center}
\bigskip
\centerline{\large\bf Abstract}
\begin{quote}\small
The primordial non-Gaussianity parameters $\fnl$ and $\tnl$ may be  scale-dependent.  
We investigate the capability of future measurements of the CMB  $\mu$-distortion, which  is very sensitive to  small scales, and of the large-scale halo bias to test 
the running of local non-Gaussianity. 
We show that, for an experiment such as PIXIE, a measurement of the   $\mu$-temperature correlation can pin down the spectral indices $n_{\fnl}$ and $n_{\tnl}$ 
to values of the order of 0.3 if $\fnl=20$ and $\tnl=5000$. A similar value can be achieved with an all-sky survey extending to redshift $z\sim 1$. 
In the particular case in which the two spectral indices are equal, as predicted in models where the cosmological perturbations are generated by a single field other 
than the inflaton, then the 1-$\sigma$ error on the scale-dependence of the non-linearity parameters goes down to 0.2.
\end{quote}

\normalsize

\newpage

\section{Introduction}
Detecting a possible primordial source of non-Gaussianity (NG) in the cosmological perturbations is one of the main targets of current and future experiments measuring the properties of the Cosmic Microwave Background (CMB) anisotropies and the large-scale structure. Indeed, measuring a certain level of NG in the three- (bispectrum) and four-point (trispectrum) correlator of the perturbations opens up a unique window into the physics
of inflation which is believed to be the period during which such fluctuations are quantum-mechanically generated \cite{reviewNG}.
The current constraints on NG come from the measurement of the CMB anisotropy bispectrum \cite{wmap} and from the properties of the clustering of galaxies which has been identified to be a  powerful probe of
NG thanks to the fact that NG introduces a scale-dependent bias between the power spectra of halos and dark matter \cite{scale,scalereview}.

Most of the attention in the literature has been devoted to the so-called ``local'' model of NG, where 
the NG is defined in terms of the primordial gravitational potential $\Phi({\vec x})$ as 

\begin{equation}
\Phi({\vec x})= \phi_{\rm G}({\vec x}) + f_{\rm NL} \Bigl[\phi^2_{\rm G}({\vec x})-\langle \phi^2_{\rm G}({\vec x}) \rangle \Bigr].
\end{equation}
The corresponding bispectrum and trispectrum are given by 
\begin{eqnarray}\label{eq:inbistris}
B_{\Phi}(k_1,k_2,k_3)&=&2f_{\rm NL}\Bigl[P_{\phi}(k_1)P_{\phi}(k_2) \mbox{ + 2 cyc}.\Bigr],\label{a1}\\
T_{\Phi}(k_1,k_2,k_3,k_4)&=&\frac{25}{9}\tau_{\rm NL}\Bigl[P_{\phi}(k_1)P_{\phi}(k_2)P_{\phi}(k_{13}) + \mbox{11 cyc}.\Bigr]\label{a2},
\end{eqnarray}
where $P_\phi(k)$ is the power spectrum of the gravitational potential. This type of NG is
generated in  multifield inflationary models where the cosmological perturbation is sourced by light scalar fields other than the inflaton. The corresponding perturbations are both scale invariant and special conformally invariant \cite{creminelli,KR}.
The parameter $\fnl$ is currently constrained to be in the range $(32\pm 21)$ by WMAP \cite{wmap} and $(28\pm23)$ by the large-scale structure \cite{slosar}, while the parameter $\tnl$ needs to be in the range $(-0.6<\tnl/10^4<3.3)$ as inferred from the WMAP 5-year data \cite{tauNL}. Measuring the amplitudes of both the bispectrum and the trispectrum is extremely interesting as,  if only one degree of freedom is responsible for the perturbations, then  there is a well-defined relation between the NG parameters, $\tau_{\rm NL}=\left(\frac{6}{5}f_{\rm NL}\right)^2$. On the contrary, if more than one field is responsible for the cosmological perturbations generated through the inflationary dynamics, then   there exists   an inequality, $\tau_{\rm NL}>\left(\frac{6}{5}f_{\rm NL}\right)^2$ \cite{SY,KR,bw}. 
To which extent future measurements of the scale-dependence of halo bias can test multi-field inequality has been the subject of Ref. \cite{biagetti}.

Even though the definitions (\ref{a1}) and  (\ref{a2})  are widely used to model NG in the primordial perturbations, it is just the first step one can make on this matter. One, more general, definition of the bispectrum and trispectrum could include a scale-dependence in the non-linearity parameters $f_{\rm NL}$ and $\tau_{\rm NL}$. This step is 
 well-motivated by the theoretical predictions of some models \cite{r1,r2,byr,r3,r4}. The   running with physical scale
of the NG parameters $\fnl$ and $\tnl$ has been the subject of an intense recent research \cite{s1,Huang1, Huang2,s2,s3,Agullo,gian,s4}.

%and, if we are in the case of a multi-field inflation, an additional scale dependence to the one coming from the spectral index in the dimensionless power spectra, as showed in \cite{byr},
%
%\begin{eqnarray}\label{eq:byr1}
%f_{\rm NL}(k_1,k_2,k_3)&=&\frac{1}{2}\frac{B(k_1,k_2,k_3)}{P(k_1)P(k_2) + \mbox{2 perm}}\nonumber\\
%\quad\nonumber\\
%		   &=&\frac{(k_1k_2)^{-3}\sum_{abcd}\mathcal{P}_{ac}(k_1)\mathcal{P}_{bd}(k_2)f_{cd}(k_3) + \mbox{2 perms}}{(k_1k_2)^{-3}\mathcal{P}(k_1)\mathcal{P}(k_2) + \mbox{2 perms}}
%\end{eqnarray}
%
%\noindent and
%
%\begin{eqnarray}\label{eq:byr2}
%\tau_{\rm NL}(k_1,k_2,k_3,k_4)&=&\frac{9}{25}\frac{T(k_1,k_2,k_3,k_4)}{P(k_1)P(k_2)P(k_{13}) + \mbox{11 perm}}\nonumber\\
%\quad\nonumber\\
%		   &=&\frac{(k_1k_2k_{13})^{-3}\sum_{abcdef}\mathcal{P}_{ac}(k_1)\mathcal{P}_{be}(k_2)\mathcal{P}_{df}(k_{13})f_{cd}(k_3)f_{ef}(k_4) + \mbox{11 perms}}{(k_1k_2k_{13})^{-3}\mathcal{P}(k_1)\mathcal{P}(k_2)\mathcal{P}(k_{13}) + \mbox{11 perms}}
%\end{eqnarray}
%
%\noindent Where we dropped the subscript $\phi$. The scale-dependent quantities $f_{ab}(k)$ and $\mathcal{P}_{ab}(k)$ are defined in eqs. (2.13) and (2.19) of \cite{byr}. 
To account for the running of $f_{\rm NL}$ in its full generality 
one can adopt for example  the parametrization  used in Ref. \cite{hut} (see also Ref. \cite{byr})

\begin{equation}\label{eq:bis}
B_\Phi(k_1,k_2,k_3)=2\left[\xi_{\fnl}(k_3)\xi_m(k_1)\xi_m(k_2)P_\phi(k_1)P_\phi(k_2) \mbox{ + cyc}.\right],
\end{equation}
where 

\begin{equation}
\xi_{\fnl,m}(k)=\xi_{\fnl,m}(k_0)\left(\frac{k}{k_0}\right)^{n_{\fnl,m}}.
\end{equation}
Here $\xi_{\fnl}(k)$ parametrizes the (self-)interactions of the fields and $\xi_{m}(k)$ the ratio of the contribution of each field.
From this general parametrization, we can also easily extend the one for the trispectrum

\begin{equation}\label{eq:tris}
T_\Phi(k_1,k_2,k_3,k_4)=\frac{25}{9}\left[\xi_{\tnl}(k_3,k_4)\xi_m(k_1)\xi_m(k_2)\xi_m(k_{13})P_\phi(k_1)P_\phi(k_2)P_\phi(k_{13}) + \mbox{cyc}.\right],
\end{equation}
where

\begin{equation}
\xi_{\tnl}(k_i,k_j)=\xi_{\tnl}(k_0)\left(\frac{k_i k_j}{k^2_0}\right)^{n_{\tnl}}.
\end{equation}
In the single-field limit, $\xi_{\tnl}(k_i,k_j)=\frac{36}{25}\xi_{\fnl}(k_i)\xi_{\fnl}(k_j)$ and $\xi_m(k)=1$. According to this parametrization, in the case of a multi-field inflation, we have three free parameters, $n_{\fnl}$, $n_m$ and $n_{\tnl}$, which describe the scale dependence of the non-linearity parameters $\fnl$ and $\tnl$ and of the dimensionless power spectra. In order to decrease the complexity of the analysis, from now on we make the assumption that $n_m$ is significantly much smaller than unity. By doing so, we are left with the following parametrization of the 
non-linear parameters

\be
\fnl(k)=\fnl^*\left(\frac{k}{k_*}\right)^{n_{f_{\rm NL}}},
\label{eq:runningfnl}
\ee
and
\be
\tnl(k_i,k_j)=\tnl^*\left(\frac{k_i k_j}{k_*^2}\right)^{n_{\tau_{\rm NL}}}.
\label{eq:runningtnl}
\ee
 CMB information alone, in the event of a significant detection of the NG component, corresponding to $\fnl = 50$ for the local model, is able to determine $n_{\fnl}$ with a 1-$\sigma$ uncertainty of about 0.1  for the Planck mission \cite{s1}. A local bias analysis performed in the same
 Ref. \cite{s1} showed that high-redshift surveys ($z > 1$) covering a large fraction of the sky
corresponding to a volume of about $100\, h^{-3}$ Gpc$^3$  might provide a 1-$\sigma$ error on the running $\fnl$ parameter of the order of  $0.4(50/\fnl)$. 
 On the other hand, 
using the WMAP temperature maps,  a constraint on the  running of the scale-dependent parameter $\fnl$ has been recently obtained in Ref. \cite{beckerhut} to be $n_{\fnl} = 0.30(+1.9)(-1.2)$ at 95\% confidence, marginalized over the amplitude $\fnl^*$. To the best of our knowledge, no forecasts for the running of the trispectrum parameter $\tnl$ exist in the literature. In fact, in the case in which the perturbations are sourced by a single field, then a well-defined relation between the running spectral indices holds,
 
\be
n_{f_{\rm NL}}= n_{\tau_{\rm NL}}
\ee
and the indices are therefore not independent. 
In this paper we will assume that
$\fnl$ and $\tnl$, and therefore their spectral indices too,  are not related to each other, thus leaving open the possibility that the perturbations are originated from a multi-field scenario.

The goal of this paper is to provide some useful forecasts on the spectral indices
$n_{f_{\rm NL}}$ and $n_{\tau_{\rm NL}}$ from the possible physical  imprints that  NG can leave on the 
 the CMB $\mu$-distortion and the halo bias. 
Measurements of the $\mu$-type distortion of the CMB spectrum provide the unique opportunity to probe these scales over the unexplored range from 50 to $10^4$ Mpc$^{-1}$ and it has been recently pointed out   that correlations between $\mu$-distortion and temperature anisotropies can be used to test Gaussianity at these very small scales. In particular the $\mu$-temperature cross correlation is proportional to the very squeezed limit of the local primordial bispectrum and hence measures $\fnl$, while the $\mu$-$\mu$ is proportional to the primordial trispectrum and measures $\tnl$  \cite{pajerzald} (see also \cite{Ganc}). Being the $\mu$-distortion localized at small scales, we expect it to be
very sensitive to the possible running of the NG parameters $\fnl$ and $\tnl$. This will be studied in section 2. In section 3 we will study
the effect of running NG parameters onto the halo bias, taking into account the running of the trispectrum amplitude as well. Our conclusions will be presented in section 4. In all illustrations, the cosmology is a flat $\Lambda$CDM Universe with normalisation 
$\sigma_8=0.803$, Hubble constant $h_0=0.701$ and matter content $\Omega_{\rm m}=0.279$.

%%%%%%%%%%%%%%%%%%%%%%%%%%%%%%%%%%%%%%%%%% MU DISTORTION %%%%%%%%%%%%%%%%%%%%%%%%%%%%%%

\section{CMB $\mu$-distortion}
The goal of this section is to compute the effect of the running NG onto the CMB $\mu$-distortion. The latter is caused by the energy injection originated by the dissipation of acoustic waves through the Silk damping as they re-enter the horizon and start oscillating. 
 The interesting property is that this effect is related to primordial perturbation scales of $50 \lesssim k\, {\rm Mpc} \lesssim 10^4$ which are not accessible from CMB anisotropies observations.

At early times ($z \gg z_{\mu,i} \equiv 2 \times 10^6$), the content of the universe can be described by a photon-baryon fluid in thermal equilibrium which has a black-body spectrum. This equilibrium is achieved mainly through elastic and double Compton scattering. However, at later times ($z_{\mu,f}\equiv 5\times10^{4} \lesssim z \lesssim z_{\mu,i}$), double Compton scattering is no longer efficient whereas the single Compton scattering still provides equilibrium. The photon number density is however frozen and only the frequency of the photons can be changed. It can be shown that any energy injection in the photon-baryon fluid will distort the spectrum by the creation of a chemical potential $\mu$. The photon number density per frequency interval is then $n(\nu)=(e^{x+\mu(x)} - 1)^{-1}$, where $x\equiv h \nu / (k_B T)$. The parameter $\mu$ due to damping of acoustic waves can then be expressed in terms of the primordial power spectrum  \cite{ch}. Using the Bose-Einstein distribution plus the
fact that the total number of photons is constant, for
an amount of energy (density) released into the plasma $\delta E/E$, one finds that $\mu\simeq 1.4 \delta E/E$, where

\begin{equation}
\frac{\delta E}{E}\simeq \frac{1}{4}\left.\langle\delta^2_\gamma({\vec x})\rangle\right|_{z_{\mu,f}}^{z_{\mu,i}},
\end{equation}
and $\langle\delta^2_\gamma({\vec x})\rangle$ represents the photon energy density fluctuation averaged over one period of the acoustic
oscillations. 
As the modes of interest re-enter the horizon during the radiation phase, one finally  finds

\begin{equation}
\mu(\vec{x})\simeq 4.6\int\frac{{\rm d}^3k_1 {\rm d}^3k_2}{(2\pi)^6}\zeta_{\vk_1}\zeta_{\vk_2} e^{i \vk_+\cdot \vx} W\left(\frac{\vk_+}{k_s}\right)
\langle\cos(k_1 r)\cos(k_2 r)\rangle_p\left[e^{-(k_1^2+k_2^2)/k_D^2},
\right]_{z_{\mu,f}}^{z_{\mu,i}}.
\end{equation}
where $\zeta(\vx)=5\Phi(\vx)/3$ describes curvature perturbations, $\vec{k}_{\pm}\equiv\vec{k}_1\pm\vec{k}_2$ and in order to 
account for the fact that the distortion  arises from a thermalization
process, one  uses a top-hat filter in real space $W(\vx)$, which
smears the dissipated energy over a volume of radius $k_{D,f}^{-1}\lesssim k_s^{-1}$, where $k_{D}(z)$ is the diffusion momentum scale

\begin{equation}
k_D(z)\simeq 4.1\cdot 10^{-6} (1+z)^{3/2} \text{Mpc}^{-1}.
\end{equation}
We proceed by computing  the correlations between $\mu$-distortion and temperature anisotropy as well as $\mu \mu$ self correlation as done in \cite{pajerzald}, but allowing for a running of $f_{\rm NL}$ and $\tau_{\rm NL}$ given by Eq. \eqref{eq:runningfnl} and \eqref{eq:runningtnl}.
The curvature perturbation bispectrum in the squeezed limit ($k_3 \ll k_1 \sim k_2$) is expressed as
\begin{equation}
\langle \zeta_{\vk_1}\zeta_{\vk_2}\zeta_{\vk_3}\rangle = (2 \pi)^3 \delta^3(\vec{k}_1+\vec{k}_2+\vec{k}_3)  \frac{12}{5} f_{\rm NL}(k_-/2)P(k_-/2)P(k_+).
\end{equation}
The temperature-$\mu$ correlation therefore reads\footnote{To compute the temperature anisotropies we adopt the same approximation as in Ref. \cite{pajerzald}, that is the Sachs-Wolfe approximation. Based on the findings in Ref. \cite{Ganc}, where the full radiation  transfer function was adopted, we expect an overall decrease of the signal-to-noise ratio of order of 40\%. Later in the text, we also point out that the change of the pivot scale amounts to corrections of the order of 30\%.}
\begin{eqnarray}
C^{\mu T}_\ell&=&-6.1 \pi \frac{9}{25} f^*_{\rm NL} b \frac{\Delta^4_\zeta(k_p)}{\ell(\ell+1)}\ln\left(\frac{k_{D,i}}{k_{D,f}}\right)\label{CmuT}\nonumber \\
&\simeq &  -2.2 \times 10^{-16}f^*_{\rm NL}\frac{b}{\ell(\ell+1)},
\end{eqnarray}
where the primordial curvature spectrum is defined by $\langle \zeta_{\vk_1}\zeta_{\vk_2}\rangle = (2 \pi)^3 \delta^3(\vec{k}_1+\vec{k})2\pi^2\Delta^2_\zeta(k_1)/k_1^3$ with $\Delta^2_\zeta(k_p)=2.4\times 10^{-9}$ at the pivot scale $k_p\equiv 0.002\, \text{Mpc}^{-1}$ \cite{wmap}. The parameter $b$ is defined by
\begin{eqnarray}
\frac{b}{\ell(\ell+1)}&\equiv & \frac{2}{\ln\left(\frac{k_{D,i}}{k_{D,f}}\right)}\int {\rm d}\ln k_+ \, j_\ell(k_+ r_\ell)^2 W\left(\frac{k_+}{k_s}\right)\nonumber \\ 
& & \times \int {\rm d}\ln k_- \left(\frac{k_-}{2k_*}\right)^{n_{f_{\rm NL}}}\frac{\Delta^2_\zeta(k_-/2)\Delta^2_\zeta(k_+)}{\Delta^4_\zeta(k_p)}\left[e^{-k_-^2 / (2k^2_D(z))}\right]^{z_{\mu,i}}_{z_{\mu,f}}.
\label{eq:b}
\end{eqnarray}
The $\mu$-distortion is created during the period between $z_{\mu,i}=2\times10^{6}$ and $z_{\mu,f}=5\times10^{4}$ which implies $k_{D,i}\simeq 11600\text{ Mpc}^{-1}$ and $k_{D,f}\simeq 46\text{ Mpc}^{-1}$.
For a weak scale dependence $\Delta^2_\zeta(k)=\Delta^2_\zeta(k_p)(k/k_p)^{n_s-1}$ we obtain
\begin{eqnarray}
b &\simeq & \frac{1}{\ln\left(\frac{k_{D,i}}{k_{D,f}}\right)} \frac{1}{n_s+n_{f_{\rm NL}}-1}\left(\frac{1}{\sqrt{2}k_p}\right)^{n_s-1}\left(\frac{1}{\sqrt{2}k_*}\right)^{n_{f_{\rm NL}}} \left[k_D(z)^{n_s+n_{f_{\rm NL}}-1}\right]^{z_{\mu,i}}_{z_{\mu,f}}.
\label{eq:bapprox}
\end{eqnarray}
If we take the same pivot for $f_{\rm NL}$ as for the primordial spectrum, $k_*=k_p$, the equation above becomes the same expression as for a constant $f_{\rm NL}=f_{\rm NL}^*$ but with a shifted spectral index $n_s$ replaced by $( n_s+n_{f_{\rm NL}})$. This shows explicitly that we recover the scale invariant result for $n_{f_{\rm NL}}=0$ and we have $b \simeq 1 + 10(n_s + n_{f_{\rm NL}} -1) $ for $(n_s + n_{f_{\rm NL}} -1) \simeq  0$. 

\begin{figure}[h]
\begin{center}
\includegraphics[width=0.5\textwidth]{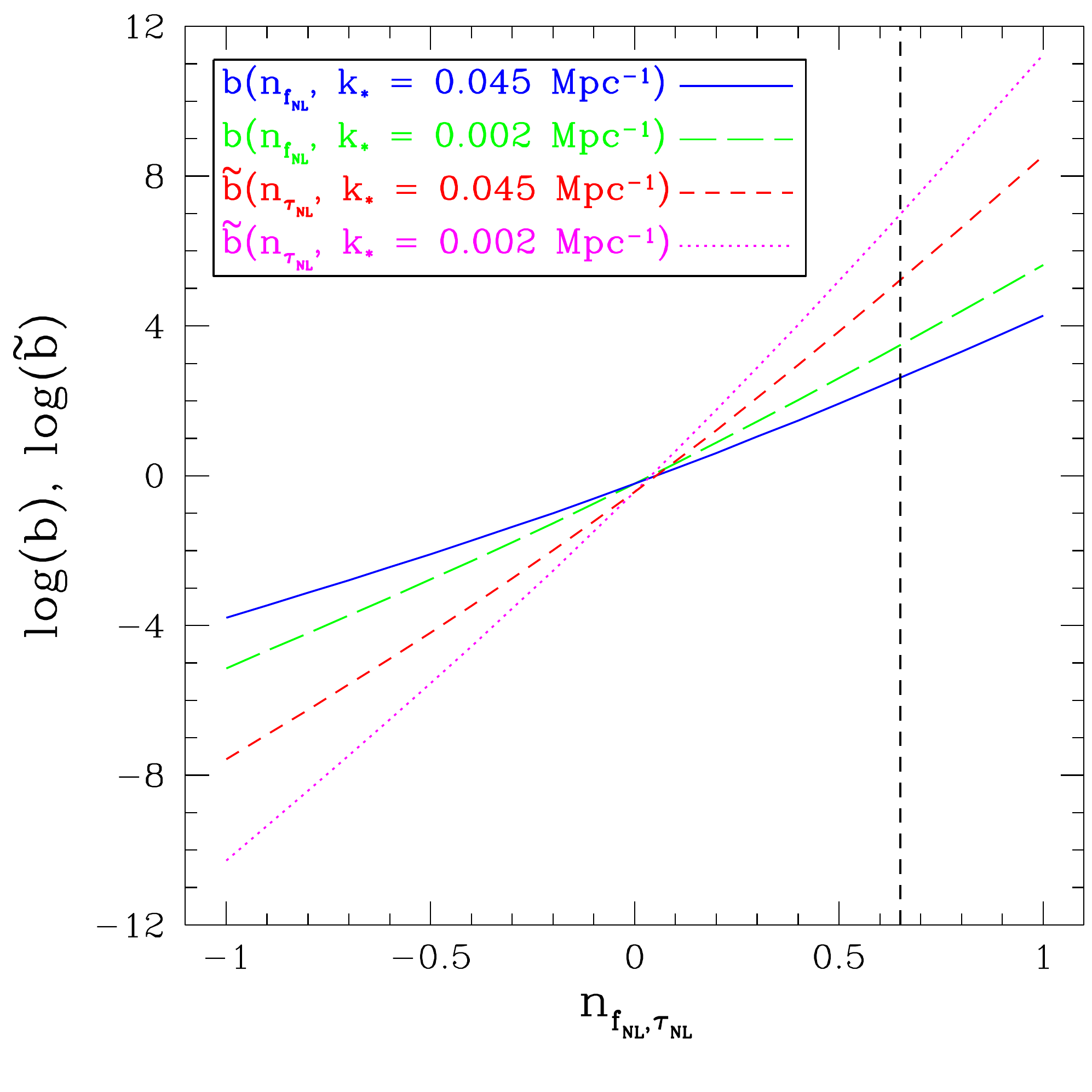}
\end{center}
\caption{\small Value of the parameters $b(n_{f_{\rm NL}})$ and $\widetilde{b}(n_{\tau_{\rm NL}})$ for two different pivot scales $k_*$ and $n_s=0.96$. The dashed line shows the maximal value of $n_{f_{\rm NL}}$ (and  $n_{\tau_{\rm NL}}$), for which the approximation \eqref{eq:sigmal} is correct.}
\label{fig:bvsn}
\end{figure}

Using the trispectrum in the collapsed limit ($\vec{k}_{12}\sim 0$)
\be 
\langle \zeta_{\vk_1}\zeta_{\vk_2}\zeta_{\vk_3}\zeta_{\vk_4}\rangle = (2 \pi)^3 \delta^3(\vec{k}_1+\vec{k}_2+\vec{k}_3+\vk_4)  4 \tau_{\rm NL}(k_-/2,k_3) P(k_-/2)P(k_+)P(k_3),
\ee
we obtain the NG contribution to the $\mu$-distortion self-correlation
\begin{eqnarray}
C^{\mu \mu}_\ell&=& 42\pi \tau^*_{\rm NL} \widetilde{b} \frac{\Delta^6_\zeta(k_p)}{\ell(\ell+1)}\ln^2\left(\frac{k_{D,i}}{k_{D,f}}\right)\nonumber \\
&\simeq & 5.6 \times 10^{-23} \tau^*_{\rm NL} \frac{\widetilde{b}}{\ell(\ell+1)},
\end{eqnarray}
where 
\begin{eqnarray}
\widetilde{b}&\equiv & \frac{2l(l+1)}{\ln^2\left(\frac{k_{D,i}}{k_{D,f}}\right)}\int {\rm d}\ln k_+  {\rm d}\ln k_-\, {\rm d}\ln k_3  \,j_\ell(k_+ r_\ell)^2 W\left(\frac{k_+}{k_s}\right)\nonumber \\ 
& & \times \left(\frac{k_- k_3}{2 k_*^2}\right)^{n_{\tau_{\rm NL}}}\frac{\Delta^2_\zeta(k_-/2)\Delta^2_\zeta(k_+)\Delta^2_\zeta(k_3)}{\Delta^6_\zeta(k_p)}\left[e^{-k_-^2 / (2k^2_D)}\right]^{z_{\mu,i}}_{z_{\mu,f}}\left[e^{-2k_3 /(2k^2_D)}\right]^{z_{\mu,i}}_{z_{\mu,f}}\nonumber \\
&\simeq &\frac{1}{\ln^2\left(\frac{k_{D,i}}{k_{D,f}}\right)} \left(\frac{1}{n_{\tau_{\rm NL}}+n_s-1}\right)^2\left(\frac{1}{\sqrt{2}k_p}\right)^{2(n_s-1)}\left(\frac{1}{\sqrt{2}k_*}\right)^{2n_{\tau_{\rm NL}}}\left(\left[k_D(z)^{n_{\tau_{\rm NL}}+n_s-1}\right]^{z_{\mu,i}}_{z_{\mu,f}}\right)^2.
\end{eqnarray}
This is just $b^2$ with the index $n_{f_{\rm NL}}$ replaced by $n_{\tau_{\rm NL}}$ and it corresponds to the result of a constant $\tau_{\rm NL}=\tau_{\rm NL}^*$ with $n_s$ replaced by $(n_s+n_{\tau_{\rm NL}})$. We recover the scale-invariant result for $n_{\tau_{\rm NL}}=0$. The behaviour of the parameters $b$ and $\widetilde{b}$ is shown on Fig.  \ref{fig:bvsn}.

%The gaussian contribution $C^{\mu \mu}_{l,Gauss}\sim 10^{-28}$ is negligible as shown in \cite{pajerzald} and the temperature self-correlation is $C^{TT}_{l}=\frac{2\pi}{25}\frac{\Delta^2_\zeta(k_p)}{l(l+1)}\simeq \frac{6.0\times 10^{-10}}{\ell(\ell+1)}$ as usual.
Having computed the key parameters $b$ and $\widetilde{b}$, we proceed by  estimating the signal-to-noise ratio to estimate
the values of $n_{f_{\rm NL}}$ and $n_{\tau_{\rm NL}}$ measurable from the $\mu$-distortion assuming that 
the amplitude $f^*_{\rm NL}$ and $\tau^*_{\rm NL}$ are known from some other experiments. In general the signal-to-noise ratio for variables $\lambda_i$ is defined in terms of the Fisher matrix as \cite{james}
\begin{equation}
\frac{S}{N}\equiv\sqrt{\lambda_i F_{ij} \lambda_j}.
\end{equation}
In the case of only one variable, it reduces to $S/N=\lambda\sqrt{F}=\lambda / \sigma_{\lambda}$.
In our case, to measure the spectral index $n_{f_{\rm NL}}$ we can adopt the Fisher matrix

\begin{equation}
F=\sum_{\ell\geq 2}\frac{1}{\sigma_{C_\ell^{\mu T}}^2} \left(\frac{\partial C_\ell^{\mu T}}{\partial n_{f_{\rm NL}}}\right)^2,\label{eq:fishermu}
\end{equation}
while for the spectral index $n_{\tau_{\rm NL}}$ we adopt the Fisher matrix
\begin{equation}
F=\sum_{\ell\geq 2}\frac{1}{\sigma_{C_\ell^{\mu \mu}}^2} \left(\frac{\partial C_\ell^{\mu \mu}}{\partial n_{\tau_{\rm NL}}}\right)^2\label{eq:fishertau}.
\end{equation}
%where ${\lambda_\alpha}$ are the parameters of our theory and where we used the $C_\ell$'s as our observables. In our case, we consider $\lambda_1=f_{\rm NL}^*$ and $\lambda_2=n_{f_{\rm NL}}$. The marginalized $1\sigma$-error on the parameter $\lambda_\alpha$ is then given by $\sigma_{\lambda_\alpha}^{(marg)}=\sqrt{(F^{-1})_{\alpha \alpha}}$ provided $F$ is non-singular whereas the $1\sigma$-error on $\lambda_\alpha$ when all other parameters are fixed is simply $\sigma_{\lambda_\alpha}=1/\sqrt{F_{\alpha \alpha}}$\cite{durrerCMB}. Note that the Fisher matrix can depend on the values of the parameters of the theory in general. The value at which we chose to evaluate the Fisher matrix is the value of the parameter that we suppose to be true in our theory. We can also make forecasts on the errors on $\tau_{\rm NL}^*$ and $n_{\tau_{\rm NL}}$ using $C^{\mu \mu}_\ell$ instead of $C_\ell^{\mu T}$.
The noise for $\mu$-distortion can be modelled assuming a Gaussian beam experiment \cite{Dodelson} by
\begin{equation}
C^{\mu \mu,N}_{\ell}\simeq w_\mu^{-1} e^{\ell^2/\ell_{\rm max}^2},
\end{equation}
where $\ell_{\rm max}$ is the maximum multipole fixed by the experiment's beam size and $w_\mu$ is the sensitivity to $\mu$. For the PIXIE experiment \cite{PIXIE}, $\ell_{\rm max}=84$ and $w_\mu^{-1/2}=\sqrt{4\pi}\times 10^{-8}$.
We also approximate the variance of the $C_\ell$'s by
\begin{eqnarray}
\sigma_{C_\ell^{\mu T}}^2 &=& \langle ( C^{\mu T}_\ell )^2 \rangle - \langle C^{\mu T}_\ell \rangle^2 \nonumber \\
&=&\frac{1}{2\ell +1}\left((C^{\mu\mu}_\ell+C^{\mu\mu,N}_\ell)(C^{TT}_\ell+C^{TT,N}_\ell)+(C_\ell^{\mu T})^2\right)\nonumber \\
&\simeq &  \frac{1}{2\ell +1} C^{TT}_\ell C_\ell^{\mu \mu, N}
\label{eq:sigmal}
\end{eqnarray}
and 
\be
\sigma_{C_\ell^{\mu \mu}}^2 \simeq  \frac{2}{2\ell +1} (C^{\mu \mu, N}_\ell)^2,
\ee
where we used that $C^{TT}_\ell \gg C^{TT,N}_\ell$, $C^{\mu\mu,N}_\ell \gg C^{\mu\mu}_\ell$ and $C^{TT}_\ell C^{\mu\mu,N}_\ell \gg (C^{\mu T}_\ell)^2.$\footnote{Using the explicit expressions above,  we find that this condition is verified provided that $
(f_{\rm NL}^*b)^2, \, \tau_{\rm NL}^*\widetilde{b} < 10^7 \ell^2$.
We consider the pivots $k_*=0.002\,{\rm Mpc}^{-1}$ and $k_*=0.064 h\,{\rm Mpc}^{-1} \simeq 0.045 \,{\rm Mpc}^{-1}$. The former corresponds to the pivot $k_p$ of the primordial spectrum and the latter to the best pivot value from \cite{beckerhut}. For $\tau^*_{\rm NL}\sim (f^*_{\rm NL})^2 \sim 10^4$ and $\ell\sim 10^2$, we find that the approximation \eqref{eq:sigmal} is valid for $n_{f_{\rm NL}},n_{\tau_{\rm NL}}\lesssim (0.65-0.85)$ depending on the pivot $k_*$, see Fig. \ref{fig:bvsn} which presents the values of $b$ and $\widetilde{b}$ as function of the indices $n_{f_{\rm NL}}$ and $n_{\tau_{\rm NL}}$ for the two pivots. One should be aware that in the multiple field case, $\tau^*_{\rm NL}$ is larger than $((6/5) f_{\rm NL}^*)^2$, so the approximation becomes  worse. In general, it seems reasonable to trust our estimation up to $n_{f_{\rm NL}}, n_{\tau_{\rm NL}}\simeq 0.5$.}

The signal-to-noise for $n_{f_{\rm NL}}$ at fixed $f_{\rm NL}^*$ is given by
\begin{eqnarray}
\left(\frac{S}{N}\right)_{n_{f_{\rm NL}}}&=&  n_{f_{\rm NL}} / \sigma_{n_{f_{\rm NL}}}(n_{f_{\rm NL}}) \nonumber \\
&=& n_{f_{\rm NL}} \sqrt{w_\mu\ln\left(\frac{l_{\rm max}} {2}\right)} 11 \sqrt{\pi} \Delta_\zeta^3(k_p)  f^*_{\rm NL} \left(\frac{1}{\sqrt{2}k_*}\right)^{n_{f_{\rm NL}}}  \left(\frac{1}{\sqrt{2}k_p}\right)^{n_s-1}\left(\frac{1}{n_{f_{\rm NL}}+n_s-1}\right) \nonumber \\
 &&\times  \left( \left(\ln \left( \frac{1}{\sqrt{2}k_*}\right)-\frac{1}{n_{f_{\rm NL}}+n_s-1}\right)\left[k_D^{n_{f_{\rm NL}}+n_s-1}\right]^{z_{\mu,i}}_{z_{\mu,f}}+\left[\ln(k_D)k_D^{n_{f_{\rm NL}}+n_s-1}\right]^{z_{\mu,i}}_{z_{\mu,f}}\right) ,
\end{eqnarray}
whereas the signal-to-noise for $n_{\tau_{\rm NL}}$ at fixed $\tau_{\rm NL}^*$ is 
\begin{eqnarray}
\left(\frac{S}{N}\right)_{n_{\tau_{\rm NL}}} &=& n_{\tau_{\rm NL}} 30 \pi w_\mu \Delta_\zeta^6(k_p) \tau^*_{\rm NL} \ln^2\left(\frac{k_{D,i}}{k_{D,f}}\right) \nonumber \\
 &&\times \left( \ln \left( \frac{1}{\sqrt{2}k_*}\right)-\frac{1}{n_{\tau_{\rm NL}}+n_s-1}+\frac{\left[\ln(k_D)k_D^{n_{\tau_{\rm NL}}+n_s-1}\right]^{z_{\mu,i}}_{z_{\mu,f}}}{\left[k_D^{n_{\tau_{\rm NL}}+n_s-1}\right]^{z_{\mu,i}}_{z_{\mu,f}}}\right) \widetilde{b}(n_{\tau_{\rm NL}},k_*) .
\end{eqnarray}
\begin{figure}[h]
\begin{center}$
\begin{array}{cc}
\includegraphics[width=0.4\textwidth]{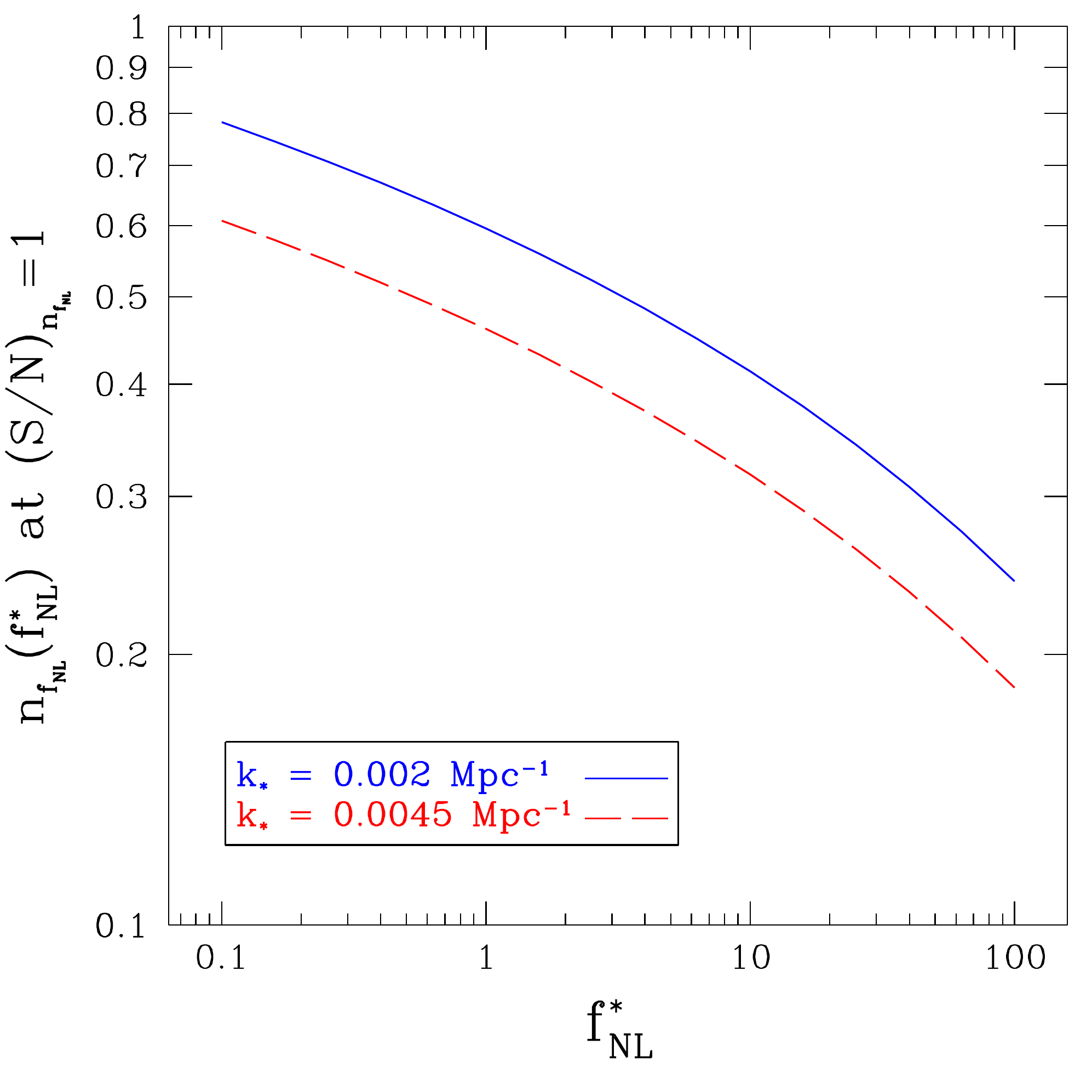} &
\includegraphics[width=0.4\textwidth]{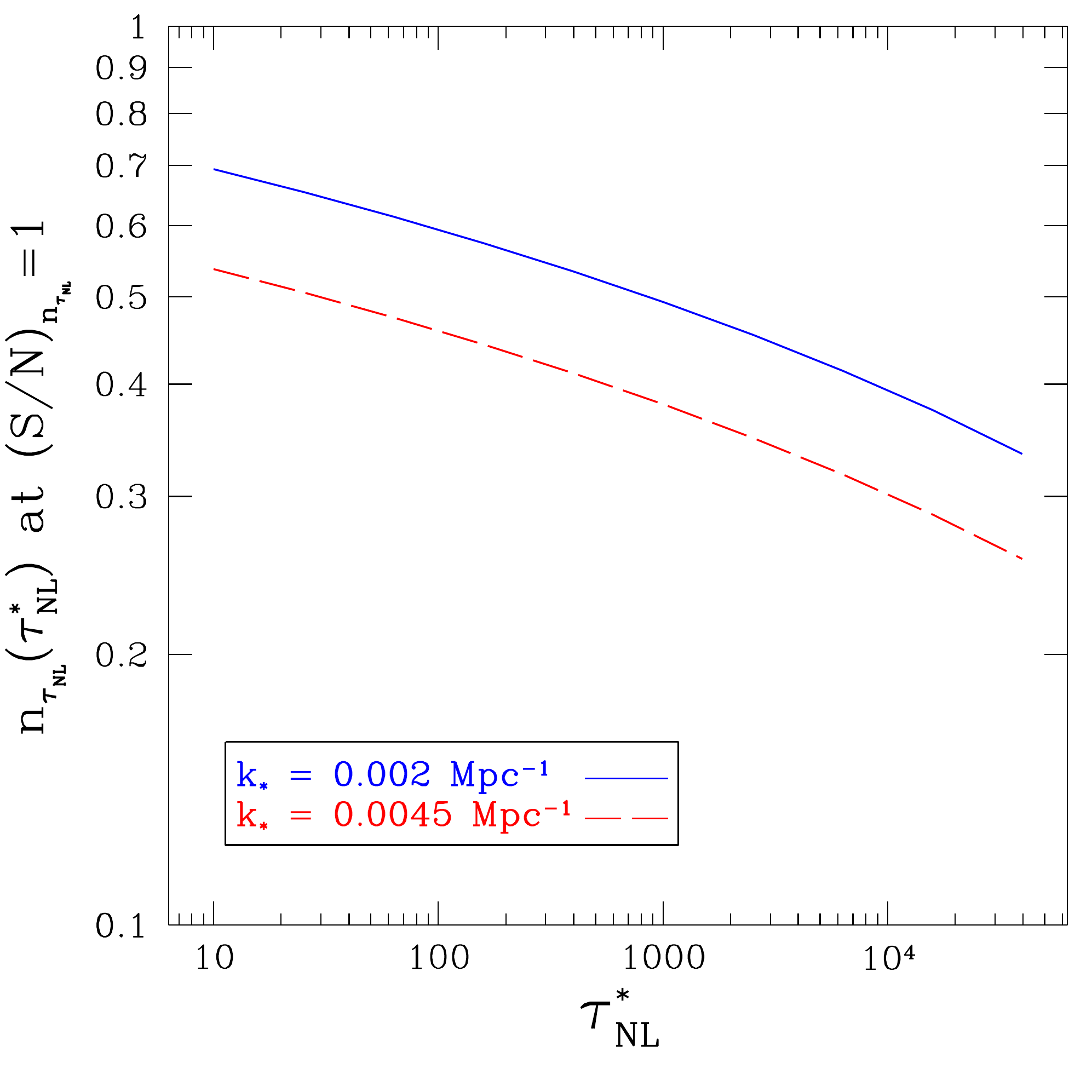} 
\end{array}$
\end{center}
\caption{Left: The spectral index $n_{f_{\rm NL}}$ as function of $f^*_{\rm NL}$ at $(S/N)_{n_{f_{\rm NL}}}=1$. Right: The  spectral index $n_{\tau_{\rm NL}}$ as function of $\tau^*_{\rm NL}$ at $(S/N)_{n_{\tau_{\rm NL}}}=1$. Both plots are made for two different pivot scales $k_*$, using $k_p=0.002\,{\rm Mpc}^{-1}$ and $n_s=0.96$ for PIXIE.}
\label{fig:SN}
\end{figure}
The left plot of Fig.  \ref{fig:SN} shows $n_{f_{\rm NL}}(f^*_{\rm NL})$ at $(S/N)_{n_{f_{\rm NL}}}=1$. 
An amplitude $f_{\rm NL}^* \lesssim 10^2$  enables to detect $n_{f_{\rm NL}} \gtrsim 0.3$, at least with the PIXIE experiment. Notice also that the  dependence on the choice of $k_*$ is relatively low.
%%The error at $n_{f_{\rm NL}}=0$ is much greater than the typical value of $n_{f_{\rm NL}}$ which could be detected if it was the true theoretical value. 
%For $f_{\rm NL}^*= 20, 50, 10^2$  and $10^3$ we have respectively $\sigma{n_{f_{\rm NL}}}(n_{f_{\rm NL}}=0)=13, 5.1, 2.6$ and 0.26 which is very high. It is therefore not possible to exclude a scale-independent non-gaussianity using this experiment alone.
The right plot of figure \ref{fig:SN} shows $n_{\tau_{\rm NL}}(\tau^*_{\rm NL})$ at $(S/N)_{n_{\tau_{\rm NL}}}=1$. Values of  $\tau_{\rm NL}^* \lesssim 10^5$ enable to detect $n_{\tau_{\rm NL}} \gtrsim 0.3$, again with the PIXIE experiment.
%The error at $n_{\tau_{\rm NL}}=0$ is much greater than the typical value of $n_{f_{\rm NL}}$ which could be detected if it was the true theoretical value. For $\tau_{\rm NL}^*= 10^4, 10^5$  and $10^6$ we have respectively $\sigma{n_{\tau_{\rm NL}}}(n_{\tau_{\rm NL}}=0)=6\times 10^4, 6\times 10^3$ and $6\times 10^2$ which is very high. It is therefore not possible to exclude a scale-independent non-gaussianity using this experiment alone.
%The dependence on $k_*$ is relatively low. The ratio of the signal-to-noise and error at $n=0$ for both $n_{\tau_{\rm NL}}$ and $n_{f_{\rm NL}}$ is about 1.3 for all $f_{\rm NL}^*$ and $\tau_{\rm NL}^*$.

In the single field case, we can use both the temperature-$\mu$-distortion correlation $C^{\mu T}$ or the $\mu$-distortion self-correlation $C^{\mu\mu}$ to measure $n_{\fnl}$. As shown in Fig.  3, $C^{\mu T}$ allows to detect lower values of $n_{\fnl}$.
\begin{figure}[h]
\begin{center}
\includegraphics[width=0.5\textwidth]{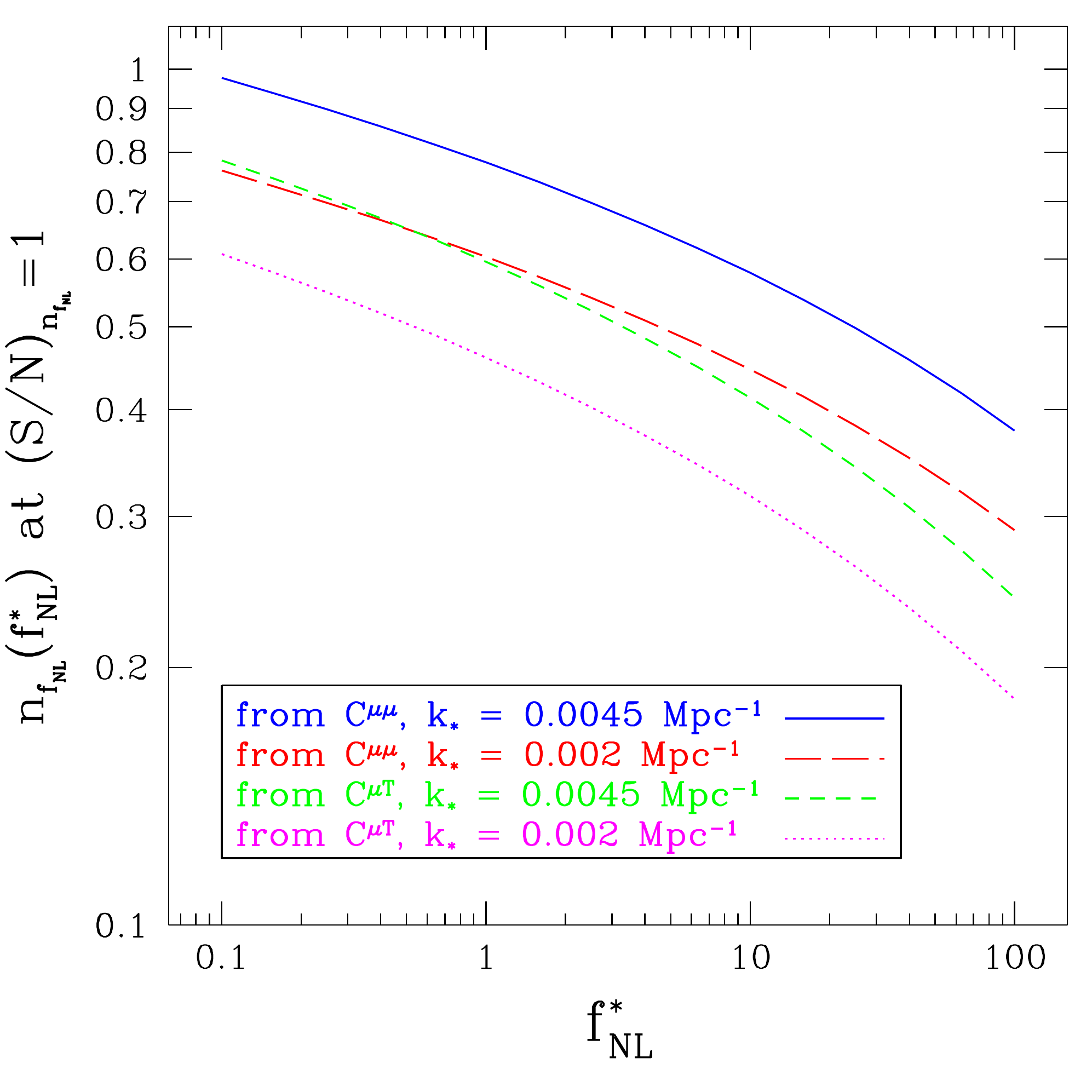}
\caption{Using the temperature-$\mu$-distortion correlation $C^{\mu T}$  allows to detect lower values of $n_{\fnl}$ than using the $\mu$-distortion self-correlation $C^{\mu\mu}$ in the single field case. The spectral index $n_{f_{\rm NL}}$ at $(S/N)_{n_{f_{\rm NL}}}$ is shown  for two different pivot scales $k_*$, using $k_p=0.002\,{\rm Mpc}^{-1}$ and $n_s=0.96$ for PIXIE.}
\end{center}
\label{fig:SNsingleee}
\end{figure}

%%%%%%%%%%%%%%%%%%%%%%%%%%%%%%%%%%%%%%%%%% HALO BIAS %%%%%%%%%%%%%%%%%%%%%%%%%%%%%%

\section{Halo bias}

Let us now turn to the effect of running NG parameters onto the halo bias \cite{s1,s2,s3,s4,hut,DJS1}.
The halo bias power spectrum with Gaussian initial conditions can be simply expressed at lowest order 
in terms of a linear (Eulerian) bias parameter
\be
P_h(k)=\left(b_1^E\right)^2 P_m(k),
\ee
where $P_m(k)$ is the dark matter power spectrum. The effect of primordial non-Gaussianity on the halo 
bias can be accurately predicted from a peak-background split \cite{slosar1,SK10,DJS1,SFL12,SH12}. As 
shown in \cite{DJS1}, the non-Gaussian contribution to the linear bias induced by a non-zero primordial
$N$-point function is
\begin{align}
\label{eq:dbk}
\Delta b_1(k) &= \frac{4}{(N-1)!} 
\frac{{\cal F}_s^{(N)}\!(k,z)}{{\cal M}_s\!(k,z)} \\
& \times \left[b_{N-2}\delta_c+b_{N-3}
\left(N-3+\frac{d\ln{\cal F}_s^{(N)}\!(k,z)}{d\ln\sigma_s}\right)\right],
\nonumber
\end{align}
where $b_N$ are Lagrangian bias parameters, $\delta_c\sim 1.68$ is the critical threshold for (spherical) 
collapse and $\sigma_s$ is the rms variance of the density field at redshift $z$ smoothed on the (small) 
scale $R_s$ of a halo. While this expression assumes a universal mass function, it can be generalized 
to take into account deviations from universality in actual halo mass functions \cite{SH12}.

The linear matter density contrast $\delta_{\vec{k}}(z)$ is related to the  curvature perturbation 
$\Phi_{\vec{k}}$ during matter domination via the Poisson equation. The latter can be expressed as 
the Fourier space relation $\delta_{\vec{k}}(z)={\cal M}(k,z)\,\Phi_{\vec{k}}$, where
\be
{\cal M}(k,z)\equiv \frac{2}{3}\frac{D(z)}{\Omega_m H_0^2}\,T(k)\,k^2. 
\ee
Here, $T(k)$ is the matter transfer function, $\Omega_m$ and $H_0$ are the matter density in critical 
units and the Hubble rate today, and $D(z)$ is the linear growth rate. 
${\cal M}_s$ is a shorthand for ${\cal M}(k,z) W(k R_s)$, where $W(k R_s)$ is a spherically symmetric 
window function (we adopt a top-hat filter throughout this paper). 
Furthermore,
\begin{align}
{\cal F}_s^{(N)}\!(k,z) &= \frac{1}{4\sigma_s^2 P_\phi(k)}
\left[\prod_{i=1}^{N-2}\int\!\!\frac{{\rm d}^3k_1}{(2\pi)^3}\,{\cal M}_s(k_i,z)\right]
{\cal M}_s(q,z) \nonumber  \\
& \quad \times \xi_\Phi^{(N)}\!(\vk_1,\cdots,\vk_{N-2},\vq,\vk)
\end{align}
is a projection factor whose $k$-dependence is dictated by the exact shape of the $N$-point function 
$\xi_\Phi^{(N)}$ of the gravitational potential. For the local constant-$\fnl$ model, the factor 
${\cal F}_s^{(3)}$ is equal to $\fnl$ in the low $k$-limit (squeezed limit), so that the logarithmic 
derivative of ${\cal F}_s^{(N)}$ with respect to the rms variance $\sigma_s$ of the small-scale density 
field vanishes on large scales. However, this does not hold for scale-dependent primordial 
non-Gaussianity. In this case, we use expressions (\ref{eq:bis}) and (\ref{eq:tris}) for the bispectrum 
and trispectrum to evaluate the derivative of ${\cal F}_s^{(N)}$ with respect to $\sigma_s$.

For generic primordial 3- and 4-point functions, the non-Gaussian halo power spectrum reads
\begin{eqnarray}\label{eq:halopow}
%P_h(k)&=& \Bigl[(b_1^E)^2+2b_1^E \Delta b_1(k)+\Delta b_{sto}(k)\Bigr]P_m(k)\nonumber\\
P_h(k) &=&\left[(b_1^E)^2+ 4 b_1^E b_1 \delta_c\frac{\mathcal{F}(n_{f_{\rm NL}},M)}{\mathcal{M}_R(k)} 
+\frac{25}{27}b_1^E\left[b_2\delta_c \sigma^2_R+b_1\left(1+\frac{d\ln{\mathcal{T}_1}}{d\ln{\sigma_R}}
\right)\right]\frac{\mathcal{T}_1(n_{\tau_{\rm NL}},M)}{\mathcal{M}_R(k)}+\right.\nonumber\\
&&\left.+\frac{25}{9} b_1^2 \delta^2_c \frac{\mathcal{T}_2(n_{\tau_{\rm NL}},M)}{\mathcal{M}^2_R(k)}
\right]P_m(k)\;,
\end{eqnarray}
where, on large scales, the last term in the square brackets can generate stochasticity between the 
halo and mass density fields if $\tnl$ is different from $(6\fnl/5)^2$ \cite{THS10,bau,GY,YM}).
We have defined the quantities
\begin{eqnarray}
\mathcal{F}(n_{f_{\rm NL}},M)&=&\frac{1}{\sigma^2_R}
\int \frac{{\rm d} q}{2\pi^2} q^2 \mathcal{M}^2_R(q)P(q)f_{\rm NL}(q),\\
\mathcal{T}_1(n_{\tau_{\rm NL}},M)&=&\frac{6}{\sigma^4_R }
\int \frac{{\rm d}^3 q_1 {\rm d}^3 q_2}{(2\pi)^6} 
\mathcal{M}_R(q_1)\mathcal{M}_R(q_2)\mathcal{M}_R(q_{12})P(q_1)P(q_2)
\tau_{\rm NL}(q_1,q_{12}),\\
\mathcal{T}_2(n_{\tau_{\rm NL}},M)&=&\frac{1}{\sigma^4_R}
\int \frac{{\rm d} q_1 {\rm d} q_2}{(2\pi^2)^2} q^2_1 q^2_2 \mathcal{M}^2_R(q_1)\mathcal{M}^2_R(q_1)
P(q_1)P(q_2)\tau_{\rm NL}(q_1,q_{2}) \;.
\end{eqnarray}
We have used the definitions (\ref{eq:runningfnl}) and (\ref{eq:runningtnl}) to obtain these
expressions. We have also emphasized the dependence on the parameters $n_{f_{\rm NL}}$ and 
$n_{\tau_{\rm NL}}$, as well as the halo mass $M$ which, for the top-hat filter, is related to the 
smoothing radius $R$ through $R=(3M/4\pi)^{1/3}$. 
The values of $\fnl^*$ and $\tnl^*$ at the pivot wavenumber $k_*=0.045$ Mpc$^{-1}$ are assumed to 
be known. 
In the particular case of scale-independent $\fnl$ and $\tnl$, i.e. $n_{f_{\rm NL}}=n_{\tau_{\rm NL}}= 0$,
we recover the expressions given in Refs.~\cite{GY} and \cite{YM}.

\vspace{2mm}

\noindent 
In order to assess the ability of forthcoming experiments to probe the scale dependence of the 
non-linearity parameters $\fnl$ and $\tnl$ through a measurement of the large scale bias, we use 
the Fisher information content on $\fnl$ and $\tnl$ (see e.g. \cite{s1,s2,s3,s4,gian,hut} for application 
to the scale-dependence of $\fnl$) in the two-point statistics of halos and dark matter in Fourier 
space. 

Computing the Fisher information requires knowledge of the covariance matrix of the halo samples,
\be\label{cov}
C_h(k,M,z) = b^2(k,M,z) P_m(k) + \frac{1}{\bar{n}},
\ee
where $\bar{n}$ is the mean number density of the survey.
In order to constrain $n_{f_{\rm NL}}$ and $n_{\tau_{\rm NL}}$, we assume that we have already 
measured $f^*_{\rm NL}$ and $\tau^*_{\rm NL}$. Moreover, since we are interested in investigating 
the possibility of a detection of the spectral indices, we take 
$n_{f_{\rm NL}}=n_{\tau_{\rm NL}}=0$ throughout as fiducial values.
The Fisher matrix is defined as follows
\begin{equation}
\mathcal{F}_{ij} = 
V_{\rm surv}\,f_{\rm sky} \int \frac{{\rm d}k\, k^2}{2\pi^2}\frac{1}{2 C^2_h} 
\frac{\partial C_h}{\partial \theta_i}
\frac{\partial C_h}{\partial \theta_j},
\label{eq:trace}
\end{equation}
where $\theta_{i}$ are the parameters whose error we wish to forecast, $V_{\rm surv}$ is the 
surveyed volume and $f_{\rm sky}$ is the fraction of the sky observed.
The integral over the momenta runs from $k_{\rm min}=2\pi/(V_{\rm surv})^{1/3}$ to 
$k_{\rm max}=0.03\, {\rm {\rm Mpc}}^{-1}/ h$, above which the non-Gaussian bias becomes smaller 
than contributions from second-order bias and nonlinear gravitational evolution. 
For illustration, we adopt the specifications of a wide-angle, high-redshift survey such as 
{\small BigBOSS} or {\small EUCLID}: $V_{\rm surv}f_{\rm sky}=50$ Gpc$^3/h^3$ at median redshift 
$z=0.7$. Furthermore, we ignore redshift evolution and assume that all the surveyed volume is at 
the median redshift.

We compute the uncertainties on $n_{f_{\rm NL}}$ and $n_{\tau_{\rm NL}}$ from a single population 
of tracers consisting of all halos of mass larger than $10^{13}M\odot/h$. 
Computing the Lagrangian bias factors from a Sheth-Tormen mass function \cite{STM} leads a 
linear and quadratic Lagrangian bias $b_1=0.7$ and $b_2=-0.4$.
We take the number density to be $\bar{n}=10^{-4}$ Mpc$^3/h^{3}$ .  

%Table \ref{tab:data} summarises the characteristics of these populations. 
%For a given mass $M$, the second-order Lagrangian bias parameter $b_2(M)$ is 
%computed from the Sheth-Tormen multiplicity function.

%\begin{table}
%\centering
%\caption{Average host halo mass, (Lagrangian) linear and quadratic 
%bias factors for the low- and high-mass halo samples at redshift $z=0.7$ used in 
%Fig.\ref{fig:ellipse}.}
%\begin{tabular}{lccc}
%$\quad$ & $\quad$ & $\quad$ & $\quad$  \\
%\toprule
%$\mbox{}$& $M (M_\odot/h)$ & $b_1$ & $b_2$ \\
%\hline\hline
%Halo 1 & $10^{12}$ & 0.2 & -0.2 \\
%Halo 2 & $10^{14}$ & 2.5 & 4.5  \\
%\bottomrule
%\end{tabular}
%\label{tab:data}
%\end{table}
%\begin{figure*}
%\center 
%\resizebox{0.40\textwidth}{!}{\includegraphics{Plots/Ellipse1sm.pdf}}
%\resizebox{0.40\textwidth}{!}{\includegraphics{Plots/Ellipse3sm.pdf}}
%\caption{Confidence ellipses obtained by the low-mass (left) and high-mass (right) sample, 
%assuming $\fnl=20$ and $\tnl= 5\times 10^3$.}
%\label{fig:ellipse1}
%\end{figure*}

%\begin{figure*}
%\center 
%\resizebox{0.40\textwidth}{!}{\includegraphics{Plots/Ellipse2sm.pdf}}
%\resizebox{0.40\textwidth}{!}{\includegraphics{Plots/Ellipse4sm.pdf}}
%\caption{Confidence ellipses obtained by the low-mass (left) and high-mass (right) sample,
%assuming $\fnl=50$ and $\tnl= 5\times 10^4$ .}
%\label{fig:ellipse2}
%\end{figure*}

\begin{figure*}
\centering
\resizebox{0.40\textwidth}{!}{\includegraphics{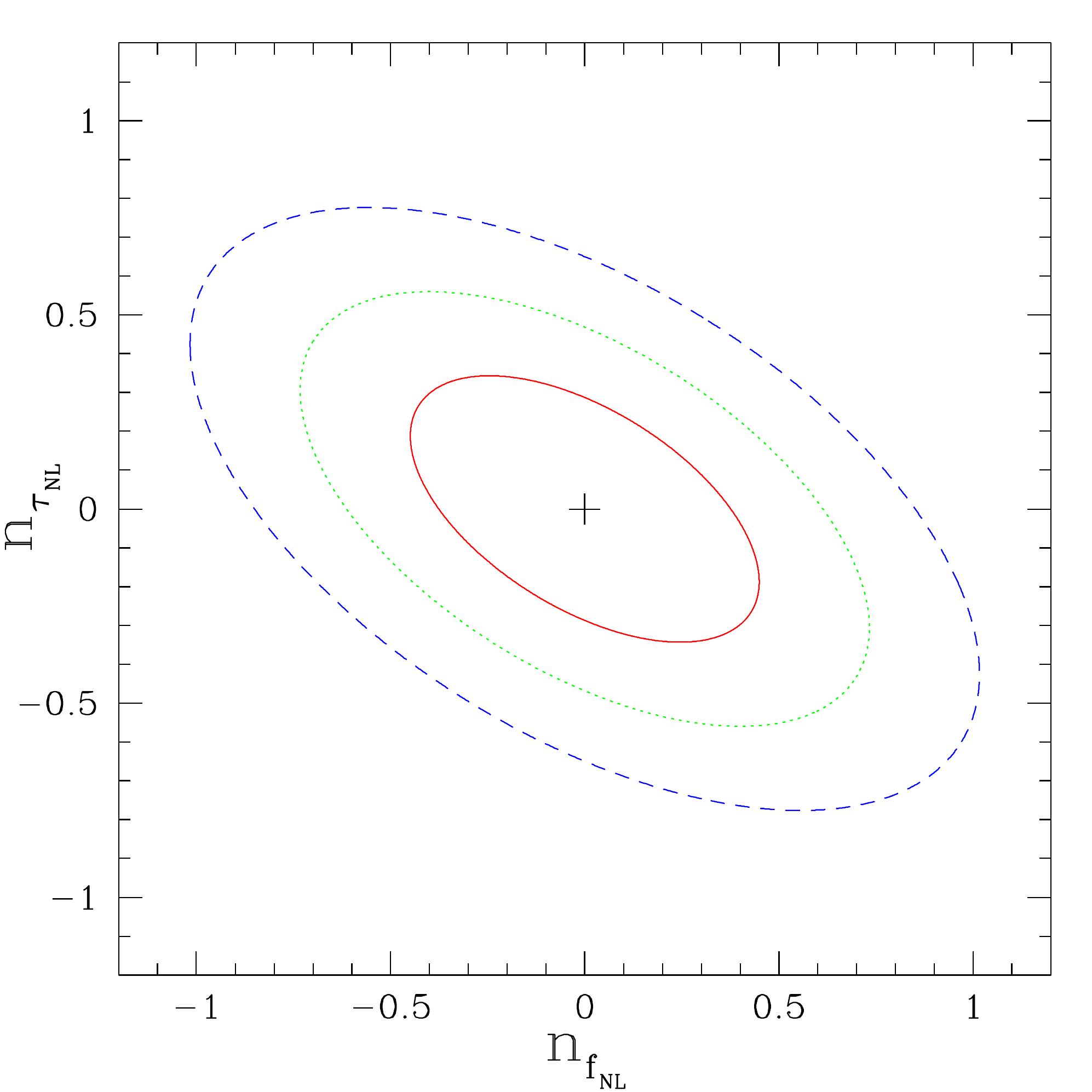}}
\resizebox{0.40\textwidth}{!}{\includegraphics{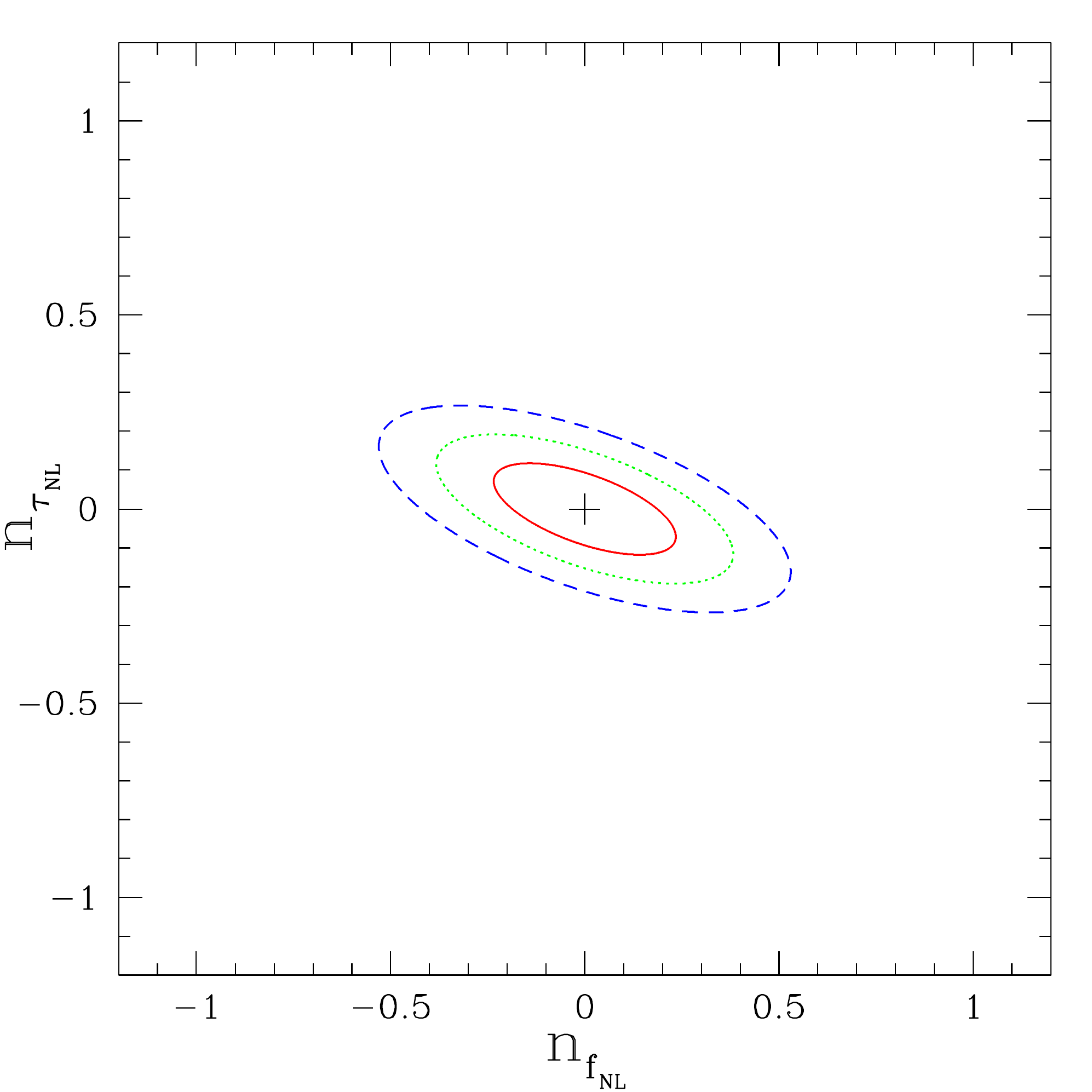}}
\caption{Confidence ellipses obtained by the population of tracers considered with halos with mass larger than $M=10^{13}M\odot/h$ ,
assuming $\fnl^*=20$ and $\tnl^*= 5\times 10^3$ (left) and $\fnl^*=50$ and $\tnl^*= 5\times 10^4$ (right).}
\label{fig:ellipse}
\end{figure*}

\begin{figure*}
\centering
\includegraphics[width=0.5\textwidth]{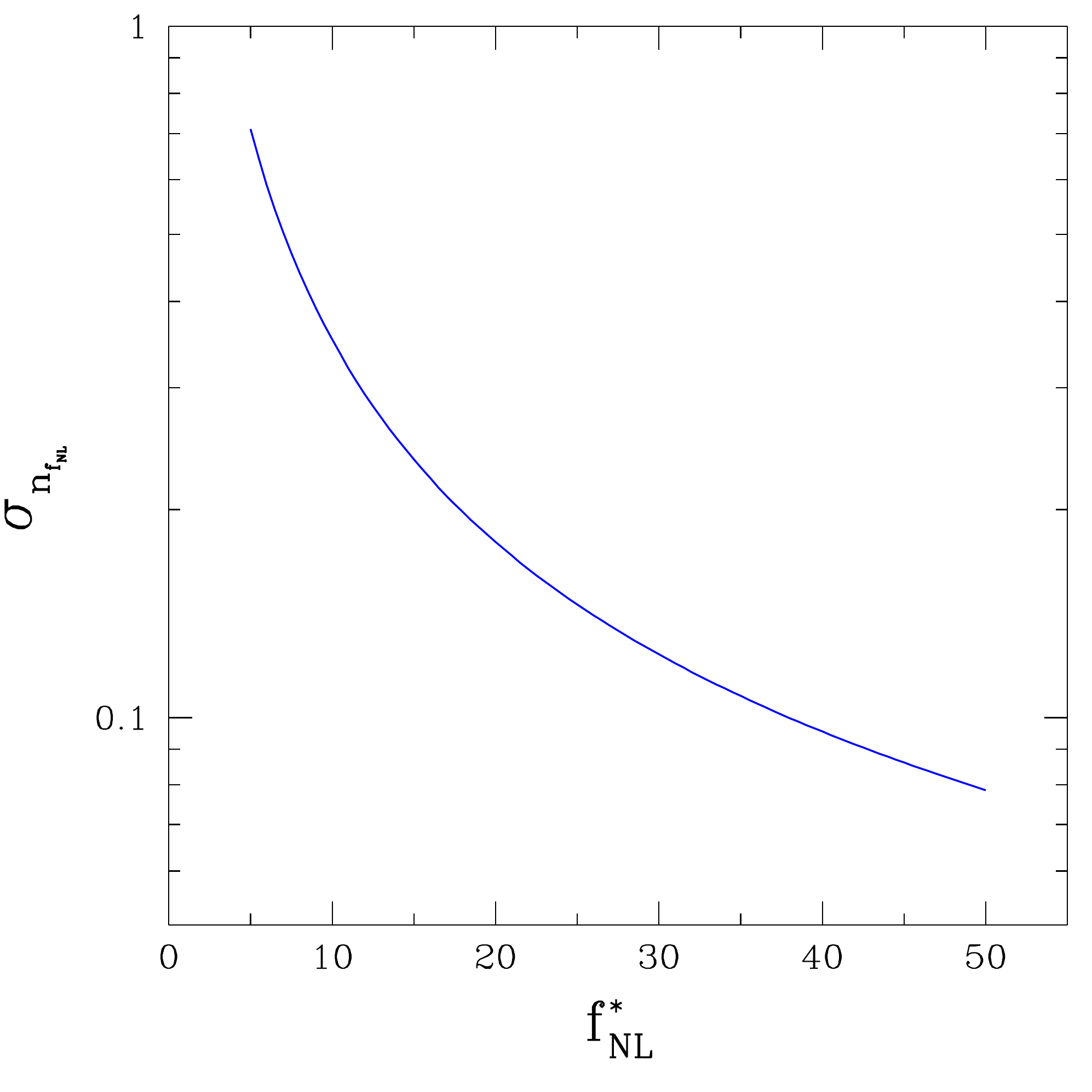}
%\resizebox{0.40\textwidth}{!}{\includegraphics{Plots/signf1.pdf}}
%\resizebox{0.40\textwidth}{!}{\includegraphics{Plots/signf2.pdf}}
\caption{1-$\sigma$ error predictions for $n_{\fnl}$ as a function of $\fnl^*$ at the pivot point $k_*=0.045 \,{\rm Mpc}^{-1}$ for the population considered in the case of one-single field models.}
\label{fig:single}
\end{figure*}

Fig. \ref{fig:ellipse} shows the resulting 68, 95 and 99\% confidence contours for the 
parameters $n_{f_{\rm NL}}$ and $n_{\tnl}$ when we assume two different combinations of 
$\fnl^*$ and $\tnl^*$. The 1-$\sigma$ errors are displayed in Table \ref{tab:data}.
In the specific case in which only one degree of freedom is responsible for the perturbations, 
we can use the relation $\tnl(k_i,k_j)=\frac{36}{25}\fnl(k_i)\fnl(k_j)$, which leaves us with 
only one parameter, $n_{\fnl}$, describing the scale dependence of the primordial NG. The 
1-$\sigma$ error for $n_{\fnl}$ as a function of $\fnl^*$ is shown in Fig. \ref{fig:single}. 
This result can be compared with those of previous work.
For a fiducial value of $\fnl^*=50$ in particular, we find an error of $\Delta n_{\fnl} \sim 0.2$ 
in the case of multi-field models, and $\Delta n_{\fnl} \sim 0.1$ in the case of single-field 
models.
For single-field models, this is a factor of ${\cal O}(3)$ lower than the forecast error found 
in Ref. \cite{s1} for a survey like {\small EUCLID}.
We attribute this difference to the fact that we have considered the higher-order term 
${\cal O}(\fnl^2)$ in the halo bias and to the parametrization 
$\fnl(K)=\fnl(k_*) (K^{1/3}/k_*)^{n_{\fnl}}$ considered in Ref.\cite{s1} for the running of 
$\fnl$. In this regards, note that $K\equiv k_1k_2k_3$ gives a contribution to the scaling of the 
external momentum, leading to a suppression (for a positive $n_{\fnl}$) or enhancement 
(for a negative $n_{\fnl}$) of the signal with respect to our parametrization in Eq. (\ref{eq:runningfnl}).\footnote{Determining $n_{f_{\rm NL}}$ through $\mu$-distortion using the parametrization of Ref. \cite{s1} also leads to a deterioration of the $S/N$ ratio. The parameter $b$ is  approximated by Eq. \eqref{eq:bapprox} with $n_{f_{\rm NL}}$ replaced by $2n_{f_{\rm NL}}/3$. The correlation $C^{\mu T}$ is decreased by a factor of about $\exp(-c\, n_{\fnl})$ with $c\simeq 3,\,4$ for $k_*=0.002, 0.045\,{\rm Mpc}^{-1}$ respectively, relative to the parametrization Eq.\eqref{eq:runningfnl}. Correspondingly,  the error $\sigma_{n_{f_{\rm NL}}}$ is increased by about $\frac{3}{2}\exp(c\, n_{\fnl})$}. 
We have checked that, if we use the parametrization and restrict ourselves to the 
${\cal O}(\fnl)$ contribution to the halo bias, we are able to reproduce their results. As noted in the introduction, the parametrization used in this paper seems to be motivated by various theoretical predictions (see for example \cite{byr,hut}).

\begin{table}\label{tab:res}
\centering
\caption{1-$\sigma$ errors for the population considered in the two different sets of $\fnl^*$ 
and $\tnl^*$ in Fig.\ref{fig:ellipse}.}
\begin{tabular}{cccc}
$\quad$ & $\quad$ & $\quad$ & $\quad$ \\
\toprule
 $\fnl^*$ & $\tnl^*$ & $\sigma_{n_{f_{\rm NL}}}$ & $\sigma_{n_{\tau_{\rm NL}}}$ \\
\hline\hline
$20$ & $5\times10^3$ & $0.30$ & $0.23$ \\
$50$ & $5\times10^4$ & $0.15$ & $0.08$ \\
\bottomrule
\end{tabular}
\label{tab:data}
\end{table}

%%%%%%%%%%%%%%%%%%%%%%%%%%%%%%%%%%%%%%%%%% JOINT FORECAST %%%%%%%%%%%%%%%%%%%%%%%%%%%%%%

%\section{Joint forecast on $f_{\rm NL}^*$ and $n_{f_{\rm NL}}$}\label{sec:joint}
%We now make a joint forecast of the errors on $f_{\rm NL}$ and $n_{f_{\rm NL}}$ if we would use both the halo bias and the $\mu$-distortion. However, the halo bias power spectrum depends two other parameter namely $\tau_{\rm NL}$ and $n_{\tau_{\rm NL}}$. \textit{We can marginalize the errors on these parameters as the errors on the experiments are uncorrelated. to check...} This can be done by the following procedure: we compute the 4x4 Fisher matrix for the halo bias and take its inverse. We remove the lines and columns corresponding to $\tau_{\rm NL}$ and $n_{\tau_{\rm NL}}$ and inverse this 2x2 matrix. This is then the marginalized Fisher matrix for the parameters $f_{\rm NL}$ and $n_{f_{\rm NL}}$ from the bias. We can now add the Fisher matrix $F_{\alpha \beta}$ from $\mu$-distortion which gives the full joint Fisher matrix $F_{tot}$. The diagonal elements of its inverse are then the marginalized errors of the parameters $f_{\rm NL}$ and $n_{f_{\rm NL}}$.
%
%Problem: the 4x4 matrix from halo bias is singular because of degeneracy problems. Dealing with it.
%
%

%%%%%%%%%%%%%%%%%%%%%%%%%%%%%%%%%%%%%%%%%% CONCLUSION %%%%%%%%%%%%%%%%%%%%%%%%%%%%%%

\section{Conclusion}
Even a tiny level of non-Gaussianity in the  cosmological perturbations can tell us a lot about 
the dynamics of the inflationary Universe. In this paper, we have focused on local non-Gaussianity,
which is a generic prediction of multifield inflationary models where cosmological perturbations 
are sourced by light scalar fields other than the inflaton. 
We have considered the possibility that the non-linear parameter $\fnl$ is scale-dependent and, 
extending the previous literature, we have also assumed that $\tnl$ may be scale-dependent. 
This is an unavoidable consequence when only a single field other than the inflaton generates 
the perturbation as the spectral indices $n_{\fnl}$ and $n_{\tnl}$ are equal. 
We have considered two possible probes of a running non-Gaussianity. First, we have exploited 
the fact that future measurements of the CMB $\mu$-distortion will be very sensitive to small 
scales, thereby enhancing the effect of a (blue) tilt of the NG parameters. 
Second, we have assessed the ability of a large-scale galaxy survey to constrain the scale 
dependence of $\fnl$ and $\tnl$ imprinted in the non-Gaussian halo bias. 
Assuming the detection of a non-vanishing $\fnl$ and $\tnl$, we find both for a CMB experiment
like PIXIE and a large-scale survey like {\small EUCLID} that the spectral indices could be measured 
with an accuracy of ${\cal O}(0.3)$ for $\fnl=20$ and $\tnl=5000$. In the case of a measurement 
of the scale-dependent halo bias, this limit could be improved by suitably combining the information from several
tracers (e.g. \cite{HSD}).

%%%%%%%%%%%%%%%%%%%%%%%%%%%%%%%%%%%%%%%%%% ACKNOWLEDGEMENT AND BIBLIOGRAPHY %%%%%%%%%%%%%%%%%%%%%%%%%%%%%%

\section*{Acknowledgments}
M.B. and V.D. are supported by the Swiss National Science Foundation (SNSF).
H.P. and A.R. are supported by the Swiss National Science Foundation (SNSF), 
project ``The non-Gaussian Universe'' (project number: 200021140236).

\begin{small}

\end{small}

\begin{thebibliography}{99}
  \bibitem{reviewNG}  For a review, see N.~Bartolo, E.~Komatsu, S.~Matarrese and A.~Riotto,
  %``Non-Gaussianity from inflation: Theory and observations,''
  Phys.\ Rept.\  {\bf 402}, 103 (2004)
  arXiv:0406398 [astro-ph.CO].
  
  \bibitem{wmap}  E.~Komatsu {\it et al.}  [WMAP Collaboration],
  %``Seven-Year Wilkinson Microwave Anisotropy Probe (WMAP) Observations: Cosmological Interpretation,''
  Astrophys.\ J.\ Suppl.\  {\bf 192}, 18 (2011)
  arXiv:1001.4538 [astro-ph.CO].
 
%  \bibitem{wmapNG} K.~M.~Smith, L.~Senatore and M.~Zaldarriaga,
%  %``Optimal limits on f_{NL}^{local} from WMAP 5-year data,''
%  JCAP {\bf 0909}, 006 (2009)
%  [arXiv:0901.2572 [astro-ph]].
   
  \bibitem{scale} N.~Dalal, O.~Dore, D.~Huterer and A.~Shirokov,
  %``The imprints of primordial non-gaussianities on large-scale structure: scale dependent bias and abundance of virialized objects,''
  Phys.\ Rev.\ D {\bf 77}, 123514 (2008)
  arXiv:0710.4560 [astro-ph.CO].
 
  \bibitem{scalereview} For a review,  see, V.~Desjacques and U.~Seljak,
  %``Primordial non-Gaussianity in the large scale structure of the Universe,''
  Adv.\ Astron.\  {\bf 2010}, 908640 (2010)
  arXiv:1006.4763 [astro-ph.CO]. 
  
  \bibitem{creminelli}P.~Creminelli,
  %``Conformal invariance of scalar perturbations in inflation,''
  Phys.\ Rev.\ D {\bf 85}, 041302 (2012)
  arXiv:1108.0874 [hep-th].
  %%CITATION = ARXIV:1108.0874;%%
  \bibitem{KR}
  A.~Kehagias and A.~Riotto,
  %``Operator Product Expansion of Inflationary Correlators and Conformal Symmetry of de Sitter,''
  Nucl.\ Phys.\ B {\bf 864}, 492 (2012)
  arXiv:1205.1523 [hep-th]; A.~Kehagias and A.~Riotto,
  %``The Four-point Correlator in Multifield Inflation, the Operator Product Expansion and the Symmetries of de Sitter,''
  Nucl.\ Phys.\ B {\bf 868}, 577 (2013)
  arXiv:1210.1918 [hep-th].
 
  
  \bibitem{slosar} A.~Slosar, C.~Hirata, U.~Seljak, S.~Ho and N.~Padmanabhan,
  %``Constraints on local primordial non-Gaussianity from large scale structure,''
  JCAP {\bf 0808}, 031 (2008)
  [arXiv:0805.3580 [astro-ph]];  N.~Afshordi and A.~J.~Tolley,
  %``Primordial non-gaussianity, statistics of collapsed objects, and the Integrated Sachs-Wolfe effect,''
  Phys.\ Rev.\ D {\bf 78}, 123507 (2008)
  arXiv:0806.1046 [astro-ph.CO].
  
  \bibitem{tauNL}  J.~Smidt, A.~Amblard, C.~T.~Byrnes, A.~Cooray, A.~Heavens and D.~Munshi,
  %``CMB Constraints on Primordial non-Gaussianity from the Bispectrum (f_{NL}) and Trispectrum (g_{NL} and \tau_{NL}) and a New Consistency Test of Single-Field    Inflation,''
  Phys.\ Rev.\ D {\bf 81}, 123007 (2010)
  arXiv:1004.1409 [astro-ph.CO].
   
  \bibitem{SY}  T.~Suyama and M.~Yamaguchi,
  %``Non-Gaussianity in the modulated reheating scenario,''
  Phys.\ Rev.\ D {\bf 77}, 023505 (2008)
  arXiv:0709.2545 [astro-ph.CO].

  
  \bibitem{bw}  G.~Tasinato, C.~T.~Byrnes, S.~Nurmi and D.~Wands,
  %``Loop corrections and a new test of inflation,''
  arXiv:1207.1772 [hep-th].
  
  \bibitem{biagetti}  M.~Biagetti, V.~Desjacques and A.~Riotto,
  %``Testing Multi-Field Inflation with Galaxy Bias,''
  arXiv:1208.1616 [astro-ph.CO], to appear in MNRAS.
  
  
  \bibitem{r1} X.~Chen,
  %``Running non-Gaussianities in DBI inflation,''
  Phys.\ Rev.\ D {\bf 72}, 123518 (2005)
  arXiv:0507053 [astro-ph.CO].
  
  \bibitem{r2}   J.~Khoury and F.~Piazza,
  %``Rapidly-Varying Speed of Sound, Scale Invariance and Non-Gaussian Signatures,''
  JCAP {\bf 0907}, 026 (2009)
  arXiv:0811.3633 [hep-th].
  
  \bibitem{byr} C.~T.~Byrnes, M.~Gerstenlauer, S.~Nurmi, G.~Tasinato and D.~Wands,
  %``Scale-dependent non-Gaussianity probes inflationary physics,''
  JCAP {\bf 1010}, 004 (2010)
  arXiv:1007.4277 [astro-ph.CO].
  
  
  \bibitem{r3} A.~Riotto and M.~S.~Sloth,
  %``Strongly Scale-dependent Non-Gaussianity,''
  Phys.\ Rev.\ D {\bf 83}, 041301 (2011)
  arXiv:1009.3020 [astro-ph.CO].
  
  \bibitem{r4}   C.~T.~Byrnes, K.~Enqvist, S.~Nurmi and T.~Takahashi,
  %``Strongly scale-dependent polyspectra from curvaton self-interactions,''
  JCAP {\bf 1111}, 011 (2011)
  arXiv:1108.2708 [astro-ph.CO].
  
  \bibitem{s1} E.~Sefusatti, M.~Liguori, A.~P.~S.~Yadav, M.~G.~Jackson and E.~Pajer,
  %``Constraining Running Non-Gaussianity,''
  JCAP {\bf 0912}, 022 (2009)
  arXiv:0906.0232 [astro-ph.CO].


  \bibitem{Huang1} 
  Q.~-G.~Huang,
  %``Negative spectral index of $f_{NL}$ in the axion-type curvaton model,''
  JCAP {\bf 1011}, 026 (2010)
  [Erratum-ibid.\  {\bf 1102}, E01 (2011)]
  arXiv:1008.2641 [astro-ph.CO].

 \bibitem{Huang2} 
  Q.~-G.~Huang,
  %``Scale dependence of $f_{NL}$ in N-flation,''
  JCAP {\bf 1012}, 017 (2010)
  arXiv:1009.3326 [astro-ph.CO].


\bibitem{s2} A.~Becker, D.~Huterer and K.~Kadota,
  %``Scale-Dependent Non-Gaussianity as a Generalization of the Local Model,''
  JCAP {\bf 1101}, 006 (2011)
  arXiv:1009.4189 [astro-ph.CO].

  \bibitem{s3}
  A.~Becker, D.~Huterer and K.~Kadota,
  %``Constraining Scale-Dependent Non-Gaussianity with Future Large-Scale Structure and the CMB,''
  arXiv:1206.6165 [astro-ph.CO].

  \bibitem{s4} M.~LoVerde, A.~Miller, S.~Shandera and L.~Verde,
  %``Effects of Scale-Dependent Non-Gaussianity on Cosmological Structures,''
  JCAP {\bf 0804}, 014 (2008)
  arXiv:0711.4126 [astro-ph.CO]. 

  \bibitem{gian}
  T. ~Giannantonio and C. ~Porciani,
  Mon.\ Not.\ Roy.\ Astron.\ Soc.\  {\bf 422} (2012) 2854-2877 
  arXiv:1109.0958 [astro-ph.CO].

  \bibitem{Agullo} 
  I.~Agullo and S.~Shandera,
  %``Large non-Gaussian Halo Bias from Single Field Inflation,''
  JCAP {\bf 1209}, 007 (2012)
  arXiv:1204.4409 [astro-ph.CO].

  \bibitem{hut} S.~Shandera, N.~Dalal and D.~Huterer,
  %``A generalized local ansatz and its effect on halo bias,''
  JCAP {\bf 1103}, 017 (2011)
  arXiv:1010.3722 [astro-ph.CO].


 \bibitem{beckerhut} A.~Becker and D.~Huterer,
  %``First constraints on the running of non-Gaussianity,''
  Phys.\ Rev.\ Lett.\  {\bf 109}, 121302 (2012)
  arXiv:1207.5788 [astro-ph.CO].
  %%CITATION = ARXIV:1207.5788;%%

  \bibitem{pajerzald} E.~Pajer and M.~Zaldarriaga,
  %``A New Window on Primordial non-Gaussianity,''
  Phys.\ Rev.\ Lett.\  {\bf 109}, 021302 (2012)
  arXiv:1201.5375 [astro-ph.CO]].

  \bibitem{Ganc} 
  J.~Ganc and E.~Komatsu,
  %``Scale-dependent bias of galaxies and mu-type distortion of the cosmic microwave background spectrum from single-field inflation with a modified initial state,''
  Phys.\ Rev.\ D {\bf 86}, 023518 (2012)
  arXiv:1204.4241 [astro-ph.CO].
 
  \bibitem{ch} J.~Chluba and R.~A.~Sunyaev,
  %``The evolution of CMB spectral distortions in the early Universe,''
  arXiv:1109.6552 [astro-ph.CO].
  %%CITATION = ARXIV:1109.6552;%%

 \bibitem{james} F. James, Statistical Methods in Experimental Physics, 2nd ed. (World Scientific, 2006).

  
  \bibitem{Dodelson}
  S. Dodelson,
  ``Modern Cosmology''
  (Academix New York, 2003).

 \bibitem{PIXIE}
 A.  Kogut, \emph{et al.},
  J.\ Cosmol. \ Astropart.\ Phys. \ 07 (2011) 025.
  
%  \bibitem{durrerCMB}
% R. Durrer,
%  ``The Cosmic Microwave Background''
%  (Cambridge University Press, 2008).
  
%  \bibitem{MV08}  S.~Matarrese and L.~Verde,
%  %``The effect of primordial non-Gaussianity on halo bias,''
%  Astrophys.\ J.\  {\bf 677}, L77 (2008)
%  arXiv:0801.4826 [astro-ph].

%  \bibitem{M08} P.~McDonald,
%  %``Primordial non-Gaussianity: large-scale structure signature in the perturbative bias model,''
%  Phys.\ Rev.\ D {\bf 78}, 123519 (2008)
%  arXiv:0806.1061 [astro-ph].


%  \bibitem{GP10} T.~Giannantonio and C.~Porciani,
%  %``Structure formation from non-Gaussian initial conditions: multivariate biasing, statistics, and comparison with N-body simulations,''
%  Phys.\ Rev.\ D {\bf 81}, 063530 (2010)
%  arXiv:0911.0017 [astro-ph.CO].

  \bibitem{slosar1} A.~Slosar,
  %``Optimal dataset combining in f_nl constraints from large scale structure,''
  JCAP {\bf 0903}, 004 (2009)
  arXiv:0808.0044 [astro-ph].


  \bibitem{SK10}  F.~Schmidt and M.~Kamionkowski,
  %``Halo Clustering with Non-Local Non-Gaussianity,''
  Phys.\ Rev.\ D {\bf 82}, 103002 (2010)
  arXiv:1008.0638 [astro-ph.CO].

  \bibitem{DJS1} V.~Desjacques, D.~Jeong and F.~Schmidt,
  %``Non-Gaussian Halo Bias Re-examined: Mass-dependent Amplitude from the Peak-Background Split and Thresholding,''
  Phys.\ Rev.\ D {\bf 84}, 063512 (2011)
  arXiv:1105.3628 [astro-ph.CO].

  \bibitem{SFL12} K.~M.~Smith, S.~Ferraro and M.~LoVerde,
  %``Halo clustering and g_{NL}-type primordial non-Gaussianity,''
  JCAP {\bf 1203}, 032 (2012)
  arXiv:1106.0503 [astro-ph.CO].

  \bibitem{SH12} R.~Scoccimarro, L.~Hui, M.~Manera and K.~C.~Chan,
  %``Large-scale Bias and Efficient Generation of Initial Conditions for Non-Local Primordial Non-Gaussianity,''
  Phys.\ Rev.\ D {\bf 85}, 083002 (2012)
  arXiv:1108.5512 [astro-ph.CO].

  \bibitem{GY}
  J.-O. Gong and S.  Yokoyama S., 2011,
  Mon.\ Not.\ Roy.\ Astron.\ Soc.\  {\bf 417}, L79
  arXiv:1106.4404 [astro-ph.CO].

  \bibitem{THS10}  D.~Tseliakhovich, C.M.~Hirata, A.~Slosar,
  %``Non-Gaussianity and large-scale structure in a two-field inflationary model''
  Phys.\ Rev.\ D {\bf 82}, 043531 (2010)
  arXiv:1004.3302 [astro-ph.CO].

  \bibitem{bau}  D.~Baumann, S.~Ferraro, D.~Green and K.~M.~Smith,
  %``Stochastic Bias from Non-Gaussian Initial Conditions,''
  arXiv:1209.2173 [astro-ph.CO].

  \bibitem{YM} S.~Yokoyama and T.~Matsubara,
  %``Scale-dependent bias with higher order primordial non-Gaussianity: Use of the Integrated Perturbation Theory,''
  arXiv:1210.2495 [astro-ph.CO].
  %%CITATION = ARXIV:1210.2495;%%

  \bibitem{STM}
  R.K.~Sheth, H.J.~Mo and G.~Tormen,
  Mon.\ Not.\ Roy.\ Astron.\ Soc.\  {\bf 323}, 1 (2001)
  arXiv:9907024 [astro-ph.CO].

  \bibitem{HSD}
  N. Hamaus, U. Seljak and V. Desjacques, 
  Phys.\ Rev.\ D {\bf 84}, 083509 (2011)
  arXiv:1104.2321 [astro-ph.CO].
  
\end{thebibliography}
\end{document}